\documentclass[12pt,preprint]{aastex}

\shorttitle{Dust Around GD362}
\shortauthors{Kilic et al.}

\begin{document}

\title{Excess Infrared Radiation from a Massive DAZ White Dwarf: GD362 -- a Debris Disk?\footnote{Based on observations obtained with the Infrared Telescope Facility, which is operated by the University of Hawaii under Cooperative Agreement no. NCC 5-538 with the National Aeronautics and Space Administration, Office of Space Science, Planetary Astronomy Program.}}

\author{Mukremin Kilic\altaffilmark{2,4}, Ted von Hippel\altaffilmark{2}, S. K. Leggett\altaffilmark{3}, and D. E. Winget\altaffilmark{2}}

\altaffiltext{2}{The University of Texas at Austin, Department of Astronomy, 1 University Station
C1400, Austin TX 78712, USA}
\altaffiltext{3}{Joint Astronomy Centre, 660 N. A'ohoku Place, University Park, Hilo HI 96720, USA}
\altaffiltext{4}{kilic@astro.as.utexas.edu}

\begin{abstract}

We report the discovery of excess K-band radiation from a massive DAZ white dwarf star, GD362. Combining
infrared photometric and spectroscopic observations, we show that the excess radiation cannot be
explained by a stellar or substellar companion, and is likely to be caused by a debris disk. This would be
only the second such system known, discovered 18 years after G29-38, the only single white dwarf currently known to be orbited
by circumstellar dust. Both of these systems favor a model with accretion from a surrounding debris disk to explain
the metal abundances observed in DAZ white dwarfs. Nevertheless, observations of more DAZs in the mid-infrared
are required to test if this model can explain all DAZs. 

\end{abstract}


\section{Introduction}

DA white dwarfs, with hydrogen-rich atmospheres, are the most common type of white dwarfs found in the Galaxy. Due to high
surface gravity, gravitational sedimentation is very fast in cool white dwarf atmospheres (Schatzman 1958).
Therefore, even if heavy elements are
initially present in the atmosphere, they would sink to the bottom of the photosphere quickly, 
leaving a pure H (or pure He for DB white dwarfs) atmosphere behind. On the other hand,
a recent survey of nearby cool DA white dwarfs by Zuckerman et al. (2003) found that up to 25\% of the white dwarfs in their sample show trace amounts
of heavy elements (mostly Ca II). Since the diffusion timescales are shorter than the evolutionary timescales, the source of
the metals observed in the photospheres of these DAZ stars cannot be primordial. Possible scenarios for the explanation
of the metal abundances include accretion from the interstellar medium (Dupuis et al. 1992), cometary impacts (Alcock et al.
1986), and accretion of asteroidal
material from a surrounding debris disk (Graham et al. 1990). Zuckerman et al. (2003) showed that the observed abundances are not compatible with
the predictions from the first two models, and they did not see any evidence for or against the third scenario.

The Sun is located in a very low density region, and the volume filling factor of clouds in the local ISM is much less than
25\%, the percentage of DA white dwarfs showing metallic features in Zuckerman et al.'s (2003) sample (see Aannestad et al. 1993). Therefore, accretion
from the ISM is unlikely to explain all of the DAZ white dwarfs. Accretion from an Oort-like comet cloud is ruled
out for two white dwarfs with high Ca abundances, G238-44 and G29-38. The Mg abundance in G238-44 was found to be stable over
the course of a year, though it should disappear in a few days if it were caused by an instantenous
cometary impact (Holberg et al. 1997). In addition, G29-38 shows significant excess infrared emission from a dust cloud (Zuckerman \& Becklin 1987), which
could not be created by a cometary impact (Zuckerman \& Reid 1998). Recently, Farihi et al. (2005) completed a survey of 371 white
dwarfs to search for low luminosity companions, and concluded that none of the white dwarfs in their sample show near
infrared excess similar to G29-38, the only single white dwarf known with a surrounding circumstellar dust disk.

Gianninas et al. (2004) reported the discovery of the most massive (1.24$M_{\odot}$) and metal-rich DAZ white dwarf ever found; GD362.
In addition to Balmer lines, they detected Ca I, Ca II, Mg I, and Fe I lines in the optical spectra of this star with
estimated $T_{\rm eff}=$ 9740 K and log $g=$ 9.1. They measured a calcium abundance of [Ca/H]$=-$5.2, about a thousand times
higher than the DAZ stars with similar effective temperatures in the Zuckerman et al. (2003) sample.
GD362, being the most metal-rich DAZ white dwarf with a Ca abundance two orders of magnitude higher than G29-38 (Koester et al. 1997),
provides a unique opportunity to test if the observed metal abundances can be explained with a debris disk similar to G29-38.
In this Letter, we present our infrared spectroscopic data for GD362.
Our observations are discussed in \S 2, while an analysis of the spectroscopic data and results from this analysis are
discussed in \S 3. 

\section{Observations}

\subsection{Photometry}

The DAZ white dwarfs identified by Zuckerman et al. (2003) are bright enough to be detected in the Two Micron All Sky Survey (2MASS).
GD362 is barely detected in the $J$-band in this survey (2MASS17313433+3705209; $J=16.16\pm0.09$) and not detected in the H and K bands.
The 2MASS Point Source Catalog provides only upper limits for the H and K band photometry of GD362. 
We derived synthetic photometry of GD362 ($J=16.04\pm0.04$, $H=16.00\pm0.03$, and $K=15.79\pm0.03$) using our infrared spectrum (see \S 2.2).
Figure 1 presents the 2MASS photometry vs. temperatures for the DAZ white dwarfs identified by Zuckerman et al. (2003), along with
the predicted sequences for DA (solid line) and DB white dwarfs (dashed line) from Bergeron et al. (1995).
The photometry from the 2MASS survey plus our synthetic photometry enable a direct comparison of the photometry for GD362
with the Zuckerman et al. (2003) sample (filled circles), including G29-38 (filled squares).
If our synthetic photometry (filled triangles) for GD362 is correct, it shows that GD362 has similar colors to G29-38.
The $J-H$ colors for G29-38 and GD362 are consistent with the model
predictions, but both objects have redder $H-K$ colors than typical white dwarfs. 

\subsection{Spectroscopy}

We used the 3m NASA Infrared Telescope Facility (IRTF) equipped with the 0.8--5.4 Micron Medium-Resolution
Spectrograph and Imager (SpeX; Rayner et al. 2003) and 0.5$\arcsec \times$15$\arcsec$ slit to obtain a resolving power of
90--210 (average resolution of 150) over the 0.8--2.5 $\mu$ range. GD362 was observed remotely from our office in Texas
under conditions of thin cirrus on
14 May 2005 starting at 13:49 UT. To remove the dark current and the sky signal from the data, the observations were taken in two different
positions on the slit (A and B) separated by 10$\arcsec$. We obtained 17 AB pairs of observations with two minute exposures
for each frame (total exposure time of 68 minutes). Internal calibration lamps (a 0.1W incandescent lamp and an Argon lamp) were used
for flat-fielding and wavelength calibration. In order to correct for the telluric features and flux calibrate the spectra,
a nearby bright A0V star (HD174567) was observed at a similar airmass with GD362. 

We note that there is a nearby star (2MASS17313587+3705357, separated by 15$\arcsec$), which is fainter than GD362 in the optical
but brighter than
GD362 in the infrared, which can cause source confusion. We obtained spectroscopy of this object as well and found it to be
an M4 star. This object is about a magnitude brighter than GD362 in the J, H, and K bands.

Here, we also present infrared observations of another white dwarf, WD2140+207, with $T_{\rm eff}=$ 8860 $\pm$ 300 K (Bergeron,
Leggett, \& Ruiz 2001), similar to GD362.
WD2140+207 and a nearby A0V star (HD210501) were observed on 19 October 2003 with the same telescope and instrument setup as GD362.
The only difference between these observations and the GD362 observations is that in 2003 we used a narrower slit (0.3$\arcsec$), which
gave an average resolving power of 250.
WD2140+207 is a cool DQ star showing molecular C$_2$ swan bands in the optical. Kilic et al. (2002) searched, without success,
for CO bands at 2.3$\mu$ in several DQ stars including WD2140+207 (Kilic et al., in preparation). DQ stars have helium rich
atmospheres, therefore have near IR spectral energy distributions similar to DB stars. Figure 1 shows that DA and DB white
dwarfs with $T_{\rm eff}>$ 5000 K differ only modestly in their IR spectral energy distributions. Hence, the WD2140+207 spectrum can be used as a spectral
template for a $\sim$9000 K DQ/DB/DA white dwarf.

We used an IDL-based package, Spextool version 3.2 (Cushing et al. 2004), designed for the reduction of spectral data obtained
with the SpeX instrument. After flat-fielding,
the spectra from individual frames were extracted using the optimal extraction
procedures described in Cushing et al. (2004). We used an extraction aperture radius of 1$\arcsec$, and each pixel in the
extraction aperture is weighted in proportion to the amount of flux it contained from the target. 

Vacca et al. (2003) presented a method for correcting the near-infrared spectra obtained with the Spex instrument for telluric
features. We use the A0V stellar spectra and a high resolution model of Vega to construct telluric spectra that are free of
stellar absorption features (see Vacca et al. 2003 for details). This method also uses the
$B$ and $V$ magnitudes for the A0V stars to redden and scale the model spectrum of Vega in order to set the flux calibration,
and therefore can be used to flux calibrate the target observations. Using the XTELLCOR package, we create a telluric spectrum
for each night, and use these spectra to correct and flux calibrate the white dwarf spectra.

\section{Results}

Figure 2 presents the extracted spectra from the flat-fielded images (no telluric correction or flux calibration)
for GD362 (darker line, top panel), WD2140+207 (darker line, bottom panel) and their reference A0V stars (red lines, scaled down to
match the white dwarf spectrum in each panel). All of the strong absorption features seen in the A0V star and
the white dwarf spectra are telluric features. A comparison of the WD2140+207 spectrum with the A0V star shows that with minor
differences in the J-band, its overall flux distribution is similar to the A0V star ($T_{\rm eff}\sim10000$K). Both GD362
and WD2140+207 have effective temperatures similar to the A0V stars,
and their flux distributions in the infrared are predicted to be similar to the Rayleigh-Jeans tail of a blackbody distribution.

The calibrated spectra for GD362 (darker line) and WD2140+207 (red line, scaled down to match the GD362 spectrum) are shown in
Figure 3 (top panel), along with the telluric and instrumental spectrum appropriate for reduction of the
GD362 observations (bottom panel).
The emission features between J \& H (1.3--1.5$\mu$) and H \& K (1.8--2.0$\mu$) bands and at the red end of
the K-band ($\geq2.5\mu$) are telluric, whereas the upward trend in the optical part of the spectrum (0.65--0.85$\mu$) is due to
the efficiency drop off of the SpeX instrument. The wavelength regions with extreme telluric features and the low
instrument-efficiency regions are shown as shaded areas in the top panel.
Dotted lines mark the location of the expected hydrogen lines (H$\alpha$ and the Paschen lines at 0.955, 1.005, 1.094, 1.282, and
1.876$\mu$) in the GD362 spectrum. The observed spectrum shows that all of these lines are detected, with the 1.282$\mu$ line
most obvious, and the 1.876$\mu$ line lost in the telluric features. Bracket lines are expected to be weaker and are only apparent in
the H band (1.5--1.8$\mu$) in the A0 star spectra (see Figure 2). These lines would be shallower in the GD362 spectra due to its
enormous gravity, and therefore would be harder to detect in our low resolution spectrum.
The 2MASS photometry for GD362 (filled circle and arrows) and G29-38 (open circles; scaled down to match the J band flux from
GD362) are also shown in Figure 3.

A comparison of the raw and the calibrated spectra for GD362 with WD2140+207 spectra (Figure 2 and 3) shows that the flux
from GD362 is consistent with a typical 9000 K white dwarf in the J and H bands, but is significantly higher than the expected
flux level from a white dwarf in the K-band. The upper limits on the H and K band photometry from the 2MASS are also consistent with
the excess K-band emission seen in the IRTF
spectrum of GD362. The amount of excess light is comparable to the amount of excess light seen from G29-38.

One caveat seen in Figure 3 is that the J-band photometry point is lower than the observed flux level from the spectra,
probably caused by the low S/N photometry from the 2MASS. Due to low resolution, it is hard to search for other
possible features in these spectra, nevertheless the Na I doublet at 2.209/2.306$\mu$ (marked with upward arrows) may be
present in the GD362 spectrum. Unfortunately, neither the spectrum of Gianninas et al. (2004) nor our spectrum covers the Na I
line at 5892\AA. On the other hand,
the spectrum of the M4 star that we observed on the same night as GD362 shows similar amount of absorption from Na I at 2.209$\mu$,
therefore the Na I feature seen in GD362's spectrum is probably real. We also note that Ca I lines at 1.927, 1.950, 1.987 and 1.992$\mu$
may be present, but they are probably lost in the telluric features.

\section{Discussion}

Tokunaga, Becklin, \& Zuckerman (1990) presented the K-band spectrum of G29-38 in their Figure 2. Their spectrum shows a constant flux
level at 6 mJy with 0.5 mJy scatter. Our IRTF spectrum of GD362 shows a constant K-band flux level at $\sim$ 0.33 mJy with
little scatter. Our spectroscopic observations and the 2MASS photometry show that GD362 has a K-band excess similar to G29-38.
Can this be another white dwarf with a surrounding debris disk?

Gianninas et al. (2004) derived an absolute V magnitude of 14.31 and a distance of $\sim$24 pc for GD362. Using these values,
we derive an apparent V magnitude of 16.21; slightly fainter than the less accurate photographic magnitude of V=16.0 (Greenstein 1980). 
The full width at half maximum of the spatial profiles in our spectral images is about 1$\arcsec$,
which corresponds to 24 AU at 24 pc. The source of the infrared excess is within 24 AU of the star.
Figure 4 shows the observed GD362 spectrum compared to a 9740 K blackbody (dashed line) normalized
to V=16.21 in each panel. The blackbody fits the observed spectrum well in the J and H bands; the distance estimate is
correct. The difference between the observed and the expected flux from the star is about 0.05 mJy in the K
band. At 24 pc, this corresponds to $M_{\rm K}\sim$ 15.8, a late type T dwarf (Leggett et al. 2002). We used a T5
dwarf template from the IRTF Spectral Library (Cushing et al. 2005) plus the normalized
9740 K blackbody to attempt to match the observed excess in the K band (red line, top panel).
T dwarfs suffer from broad water absorption bands (stronger for later spectral types) centered at 1.4, 1.9, and 2.7$\mu$.
Adding a T5 dwarf to a 9740 K blackbody creates spectral features from 1.3 to 2.5 $\mu$ that are prominent,
yet not seen in GD362's spectrum. Therefore, neither T dwarfs nor any other dwarf stars can explain the excess seen between
2.0 and 2.5 $\mu$. We also tried to match the observed spectrum with a combination of different blackbodies.
The bottom panel in Figure 4 shows the expected spectral energy distributions of a 9740 K blackbody plus 600 K, 700 K,
and 800 K blackbodies (from top to bottom). The excess emission is best fit with a 700 K blackbody.
Therefore, the best explanation for the K-band excess in GD362 is a circumstellar dust disk heated by the white dwarf.
It took 17 years to find the second white dwarf with a brown dwarf companion (Farihi \& Christopher 2004). Likewise,
our discovery of a debris disk around GD362 came 18 years after the first discovery of a debris disk around G29-38
(Zuckerman \& Becklin 1987).

Spectroscopic observations of G29-38 with the Spitzer Space Telescope revealed silicate dust around the star (Reach et al.,
in preparation), which means that the white dwarf is not the source of the dust. The observed metal abundances in GD362 cannot
be explained with the
interstellar accretion model since GD362 is well within the Local Bubble and away from any high concentration of interstellar matter
(Gianninas et al. 2004). Therefore, both G29-38 and GD362 observations favor the third model suggested by Zuckerman et al.
(2003); accretion of asteroidal material from a surrounding debris disk. If GD362 is a fainter clone of G29-38, then we
predict its flux in the 4.5$\mu$, 8$\mu$, and 24$\mu$ Spitzer bands to be 0.6 mJy, 0.5 mJy, and 0.1 mJy, respectively.
Mid-infrared photometry of a sample of DAZ white dwarfs will be necessary to check if this model can explain the observed metal
abundances in all DAZs. It appears that white dwarf observations in the infrared may have a tremendous amount to tell us about
disk and possibly planet formation and evolution.
\acknowledgements
We thank the IRTF staff, especially our telescope operators David Griep and Bill Golisch, for helping us observe remotely. 
We thank our anonymous referee for helpful suggestions that substantially improved the article.
This material is based upon work supported by the National Science Foundation under Grant AST-0307315.
This publication makes use of data products from the Two Micron All Sky Survey, which is a joint project of the University of Massachusetts and the Infrared Processing and Analysis Center/California Institute of Technology, funded by the National Aeronautics and Space Administration and the National Science Foundation.

\clearpage
\begin{figure}
\plotone{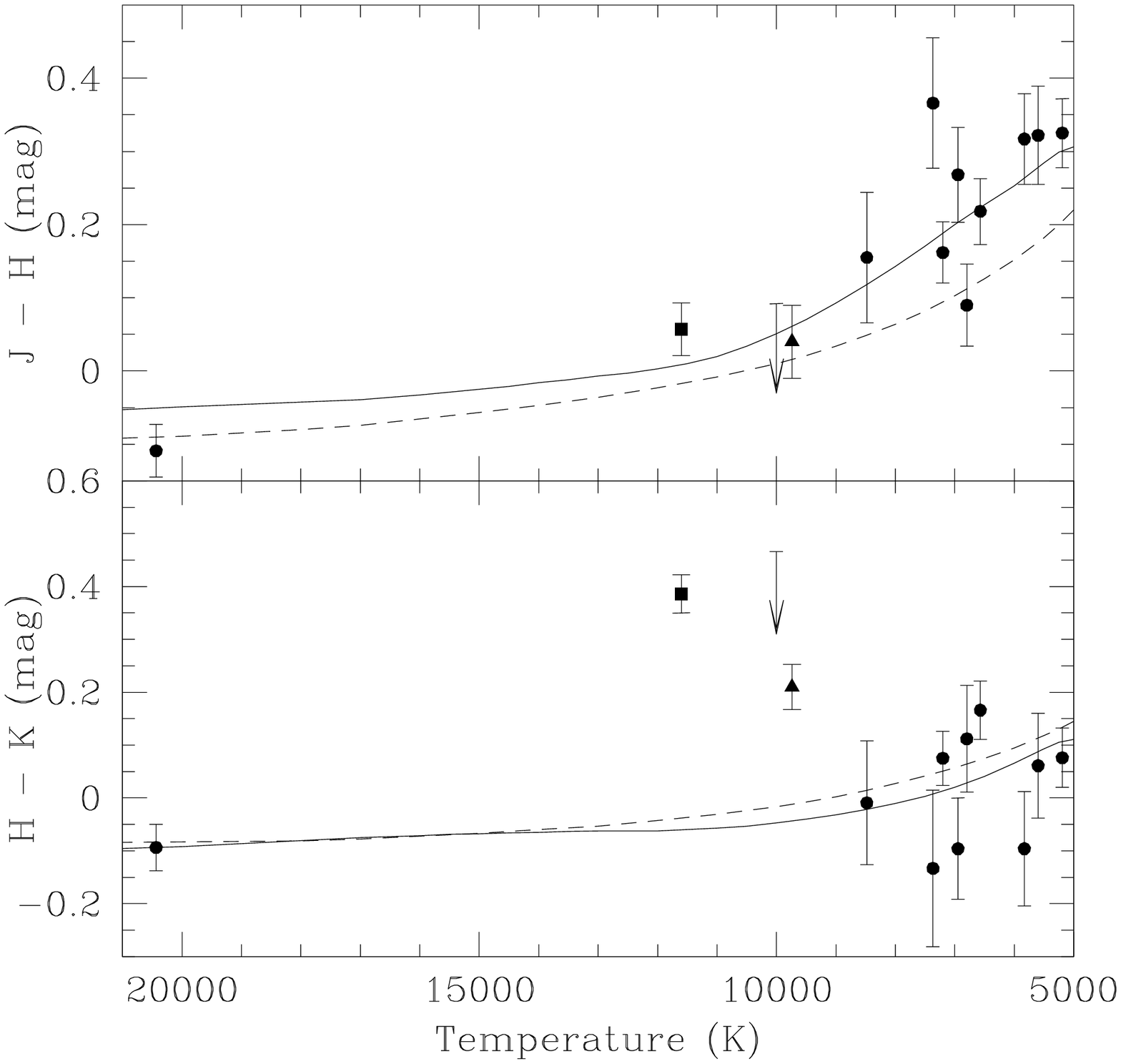}
\caption{$J-H$ and $H-K$ (from 2MASS) vs. temperature diagrams for DAZ white dwarfs identified by Zuckerman et al. (2003;
filled circles). The predicted sequences for DA (solid line) and DB (dashed line) white dwarfs and the colors for G29-38
(filled squares) are also shown. Synthetic colors derived from the IRTF spectrum of GD362 are shown as filled triangles, and these
are consistent with the 2MASS upper limits for GD362 (arrows, offset slightly in $T_{\rm eff}$ for plotting clarity).} 
\end{figure}

\clearpage
\begin{figure}
\plotone{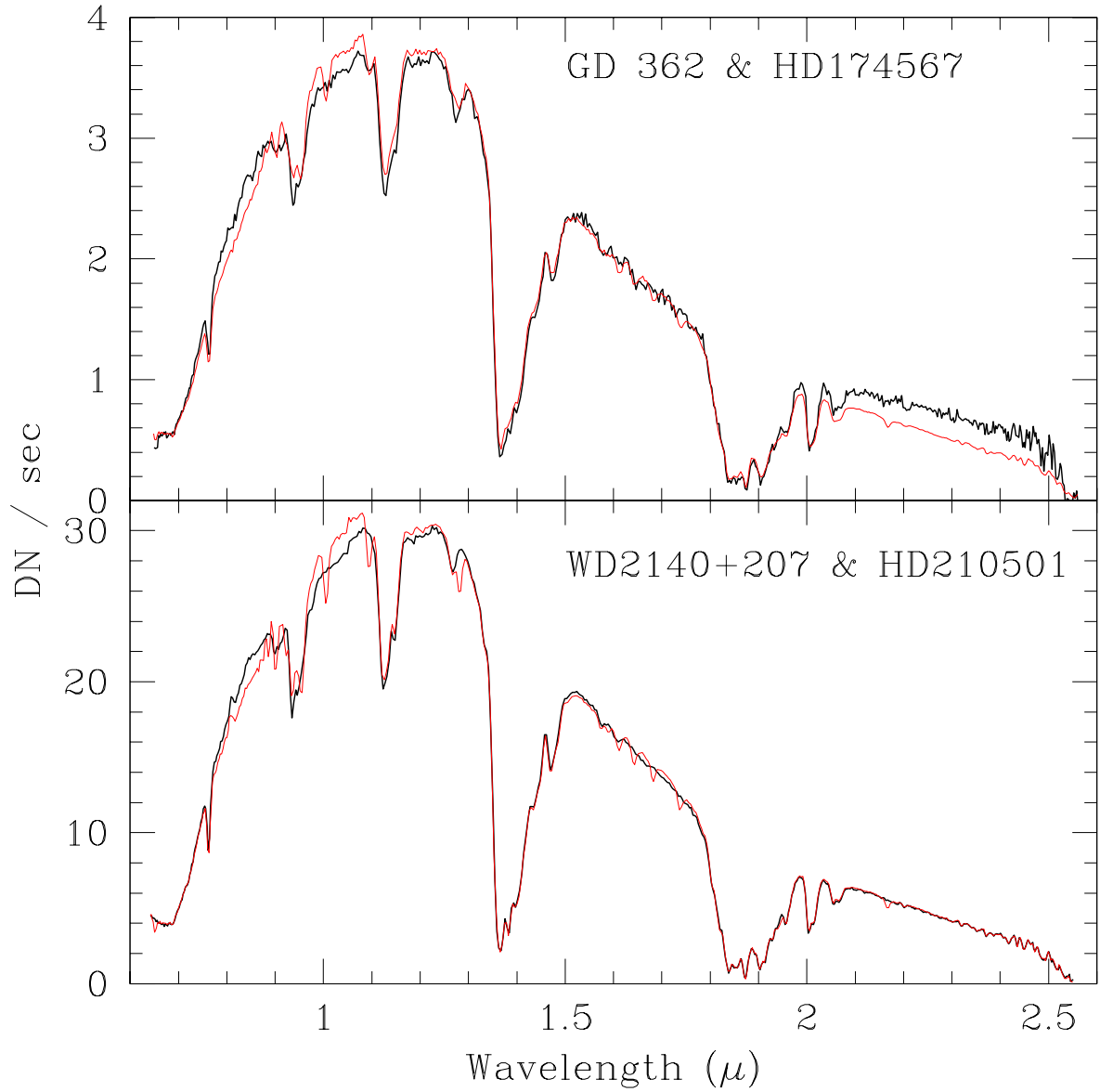}
\caption{Uncalibrated spectra of GD362 and its reference A0V star (top panel), and WD2140+207 and its reference A0V star (bottom panel).
The spectra for white dwarfs are shown in black, whereas the spectra for A0V stars are shown in red.}
\end{figure}

\clearpage
\begin{figure}
\plotone{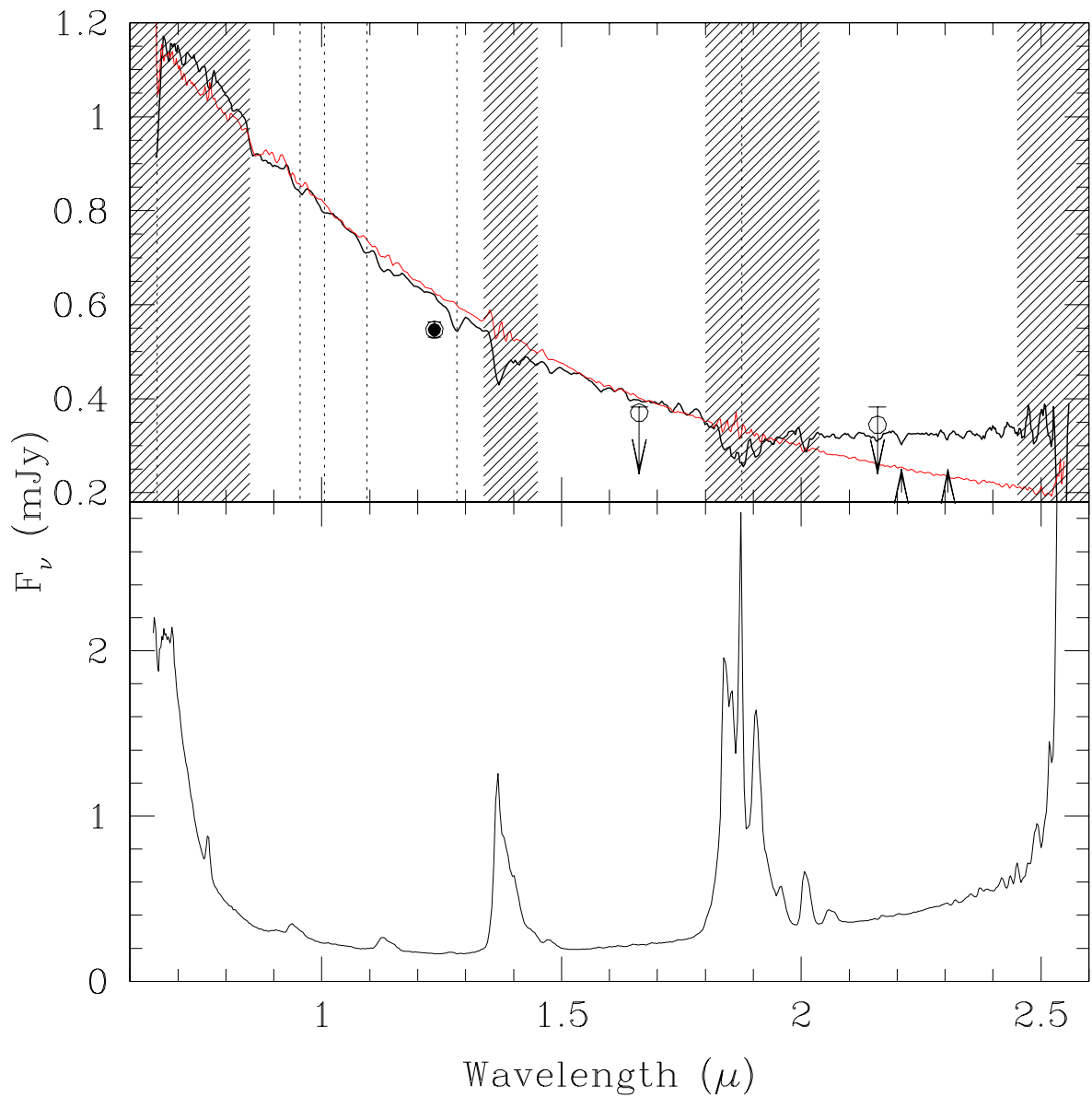}
\caption{Flux calibrated spectra of GD362 (black line) and WD2140+207 (red line), and the telluric and instrumental
spectrum appropriate for reduction of the GD362 spectrum (bottom panel).
Shaded regions in the top panel suffer from significant telluric features or instrumental efficiency losses. The 2MASS photometry for
GD362 (filled circle and downward arrows) and G29-38 (open circles, scaled down to match the J-band flux from GD362) are also shown. The dotted lines
mark the positions of H$\alpha$ and Paschen lines, whereas the upward arrows point to the expected positions of the Na I doublet.}
\end{figure}

\clearpage
\begin{figure}
\plotone{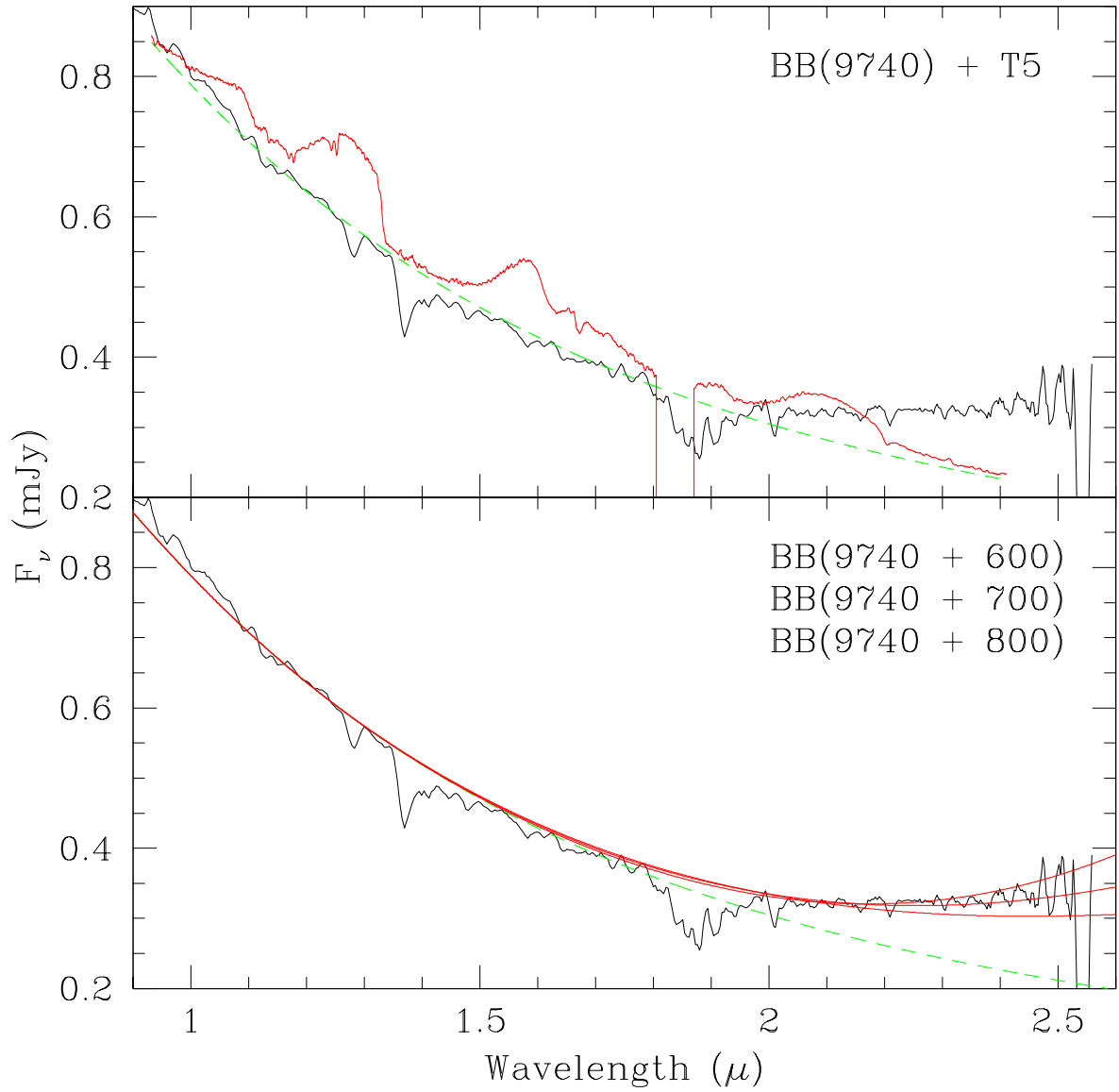}
\caption{GD362 spectrum (black line) and the expected infrared flux from a blackbody at 24 pc (dashed line). Composite blackbody +
brown dwarf/blackbody templates are shown as red lines in both panels. The bottom panel shows the effect of changing the temperature of
the secondary blackbody from 600 K to 800 K (from top to bottom).}
\end{figure}

\end{document}